# Special Purpose Computers for Statistical Physics: achievements and lessons

Lev N. Shchur
Landau Institute for Theoretical Physics, Chernogolovka, Russia;
Federal Research Center "Computer Science and Control" Russian Academy of
Sciences, Moscow, Russia;
National Research University Higher School of Economics, Moscow Russia,
lev@landau.ac.ru
ORCID: 0000-0002-4191-1324

*Abstract*— **In the late 80s and 90s, theoretical physicists of the Landau Institute for Theoretical Physics designed and developed several specialized computers for challenging computational problems in the physics of phase transitions. These computers did not have a central processing unit. They optimize algorithms to handle elementary operations on integers - read, write, compare, and count. The approach allowed them to achieve recording run times. Computers performed calculations three orders of magnitude faster than similar calculations on the world's best supercomputers. The approach made it possible to obtain fundamentally new results, some of which have not yet been surpassed in the accuracy of calculations. The report will present the main ideas for the development of specialized computers and the scientific results obtained with their help. The lessons of planning and execution of long-term complex scientific projects will also be discussed.**

## I. INTRODUCTION

The subject of statistical physics is, in particular, the study of phase transitions and critical phenomena [1]. Each of us almost daily observes one of these phenomena - the boiling of water in a kettle or coffee maker, in which the heated water, reaching the boiling point, begins to turn into steam. The phenomenon is the so-called first-order phase transition, which is characterized by the presence of latent heat, the difference between the internal energies of the two phases of water, liquid, and gaseous. Internal energy is the first derivative of the thermodynamic functional to temperature - free energy, which is what gave rise to the name phase transition of the first order at the suggestion of P. Ehrenfest [2]. The theory of first-order phase transitions has not yet been developed. Another type of phase transition is a second-order phase transition, in which the first derivative is continuous. However, there is a singularity in the second derivative, for example, the heat capacity. The highlight of the second-order phase transition is in the appearance of spontaneous magnetization of a ferromagnet when it is cooled below the critical temperature, the Curie temperature. An example of a ferromagnet is a magnetic needle, the spontaneous magnetic moment of which at room temperature acquires an orientation along the lines of force of the Earth's magnetic field. When the magnetic needle is heated above the Curie temperature, the arrow will lose its magnetic moment and, accordingly, will lose a specific orientation, and upon cooling below the Curie temperature, the orientation will be restored.

The theory of phase transitions of the second kind was built in the 40s-70s of the XX century [1]. During these years, several two-dimensional models were exactly solved [3]. Despite the tremendous theoretical advances, there are still no exact solutions to three-dimensional models. Also, the theory of systems with impurities has not been fully developed, which is very necessary for applications. It is because the chemically pure and ideally ordered systems without defects are not so widespread in real life, in practical applications and engineering solutions. The theory of glass systems, which are obtained, for example, by rapid cooling, and such materials are widespread, is also not fully developed. And it is here that the possibility and necessity of using a numerical experiment opens up.

Numerical experiments in the field of statistical physics are based on Monte Carlo methods [4]. Their use requires serious computing resources. In this post, we provide an overview of research that has been carried out using specialized computers. The main idea was to design and manufacture a computer in which the algorithm would be implemented in hardware. Such a device, in some cases, can achieve a computational speed 1000 times higher than the computation speed using the most powerful supercomputer.

The creation of such computers required a rather severe investment of time for highly qualified specialists, and without the right to make mistakes. The basis of such a project is programming using microcircuits, where it is almost impossible to fix a code error. The risk of not completing such a project is very high.

## II. SPIN-GLASS COMPUTER AT BELL LAB

Understanding the physical processes of the glass phase of a substance is one of the most critical tasks of modern solid-state physics and materials science. The study of such systems requires a lot of computer time. In the 80s of the last century, ideas for the development of specialized computers for solving a specific class of problems were implemented simultaneously in several research centers.

Practically simultaneously, processors were built to study the simplest ferromagnet model, the Ising model, at Delft Technical University [5] and the Institute for Theoretical Physics at the University of California at Santa Barbara [6]. Both approaches used the Metropolis method [7] to thermalize the system and a shift register type random number generator [8] to implement the Markov chain. In the next section, we will discuss in detail the advantages and disadvantages of these two methods, which we have discovered in our experience during the construction of specialized computers and their use for simulation. Here, we note that no new scientific results were obtained using computers in Delft and Santa Barbara.

Significant results were obtained using a computer built at Bell Laboratories [9]. We will dwell on the discussion of its design in detail. A specialized computer was designed to study a model of statistical mechanics - an Ising spin glass on a hypercubic lattice with periodic boundary conditions. At each vertex of such a lattice, spin may be absent or present (a variable with values +1 and -1). At each lattice site, there can also be an external magnetic field of a random value. Variables are located on the lattice edges, which can take on some random values. A Markov process was used to simulate the dynamics of spin configurations. The computation cycle consisted of two stages, lattice spin flips by the heat bath method [4] and the calculation of thermodynamic quantities. The first stage was performed on a dedicated processor, while the second - on a standard commercial site with a Motorola 68000 processor, which was connected to the control computer via an RS-232 serial port at a speed of 9.2 kilobits per second. This decision was made to provide flexibility in the study of models. As it turned out, this combined solution turned out to be superfluous. On the other hand, this design allowed the use of many ready-made hardware solutions for various parts and accelerated the development of the system as a whole.

The solution was chosen as follows - building a specialized high-speed processor with the implementation of one-spin relaxation and using the bus to quickly transfer large words containing packed spins to a regular computer. VMEbus with 32-bit transmission and having a separate address bus was chosen. Eight 2MB VME DRAM modules with 120 ns read time, and 240 ns memory cycle was installed.

The specialized processor was built on TTL technology and was placed on two VME cards. To change the parameters of the model was required to manufacture and install a new PROM memory and PALS (Programmable Array Logic Devices) in a specialized processor. We used the packing of spins (one-bit variable) into words. The packing, in turn, required the development of a special device for calculating the address of adjacent spins. Implementing the Markov process, pipelined processing was used. To store the probabilities of spin flips, we used a pre-calculated and loaded table implemented in 25 ns memory. A generator at a frequency of 25 MHz was used for synchronization. In the random number generator, the Fibonacci algorithm was implemented with a characteristic pair of constants (5,17).

As a result, a simulation rate of up to 17 million spin flips per second was achieved with a theoretical speed of 25 million per second. The memory capacity made it possible to study square lattices with a side of 8192, cubic with a side of 512, and four-dimensional with a side of 64, that is, 64 million spins. For comparison, the speed of calculating a similar problem on the Cray-1 supercomputer was ten times slower, depending on the type of problem. The undoubted advantage of a specialized computer was its low cost and availability of computing time.

Investigating only one problem, a three-dimensional Ising spin glass, took one year of continuous operation of a specialized computer system. New and unexpected scientific results were obtained [10,11]. A phase transition to the glassy phase at a temperature of 1.2 was discovered, and the relaxation properties of this model were studied. Note that the linear size of the lattice in the calculations did not exceed 64. That is, only 1/64 of the installed memory was used.

Summing up, let us note that excesses were introduced into the design: 1) 16 megabytes of memory, although no more than 0.5 megabytes were actually used, 2) a random external field was not used, 3) the randomness of the lattice was not used, 4) only one was investigated a task that did not require reinstalling the PROM.

The design flaw was also the slowness of processing averages on the Motorolla 68000 processor, as well as the weak properties of the random number generator for this amount of computation.

Nevertheless, we emphasize that the three-dimensional model of spin glass has not yet been studied with such accuracy by other authors, which is an undoubted scientific success of the project.

### III. SPECIAL PURPOSE PROCESSORS FOR DILUTED TWO-DIMENSIONAL ISING MODEL

The Ising model with impurities was analytically investigated by the Dotsenko brothers and required independent confirmation of their results. For this purpose, the first specialized processor was built [12]. It was the ideal Monte Carlo computer in terms of architecture. All operations were completed in less than 250 ns, with speed determined by a 200 ns memory chip cycle and 90 ns access time. As a result, the processor was performing at least 4 million operations per second—precisely operations, not just spin flips. The processor consisted of memory, logic, and counters. During the same time, the average values of the thermodynamic values over the lattice (energy and magnetic moment) were calculated using 34-bit counters, the values of which were then read in parallel with the calculations by the AT-286 control computer. The latter also counted the moments of these quantities and an estimate of the heat capacity and magnetic susceptibility. The linear size of the lattice was limited to 256 nodes. The random number generator was built on the basis of a shift register with a characteristic pair (147,250). An algorithm previously tested on a VAX-11/780 computer was implemented. The specialized processor performed only integer operations. In this aspect, the implementation of some of the operations was closer to the implementation of the Bell Lab processor [9].

After a long time of work, the operator's methodical oversight was unexpectedly revealed. The random number generator was initialized with the same sequence of 250 numbers, which led to the absolute identity of the Markov process implementations. By that time, it became clear that the lattice size of this processor would not allow revealing the subtle properties of the impurity Ising model, despite the fact that our SPP-1 processor was more flexible in this respect than its predecessors and allowed using any one-spin algorithm.

Fortunately, during the creation of the SPP-1 processor, cluster algorithms were developed for spin models, and it became clear that the simplest and most efficient of them can also be implemented in the integer arithmetic of a specialized processor. The first cluster Monte Carlo processor, SPP-2 [13], was built, which implements the one-cluster Wolff algorithm [14] for the two-dimensional Ising model with random connections. SPP-2 could be programmed to study lattices with linear dimensions in powers of two from 64 to 1024 and with a programmable random number generator such as a shift register with a length of no more than 255. The combination of a single-cluster algorithm and a random number generator was considered the best in the scientific community at that time.

Analysis of the simulation results indicated that the estimates obtained for lattice sizes of 128 and less are not accurate enough and, possibly, contain a systematic error of an unknown nature. At the same time, an article [15] appeared, which indicated that the combination of the two best methods, the single-cluster algorithm and the shift register, leads to large systematic errors. Fortunately, we managed to find a solution to this problem - a theory of the algorithm was built with the simultaneous use of a single-cluster algorithm and a shift register [16-18] and universal boundaries were established for the level of systematic errors. It turned out that, within the limits of statistical errors, we have the right to use only gratings with a linear size of 256 or more. All this allowed us to confidently publish the results for such lattice sizes. We were able to calculate for the first time the correlation function of the Ising model at the transition point and discover its interesting properties for the impurity model [19], as well as accurately estimate the critical properties of the impurity Ising model [20], which made it possible to unambiguously single out correct analytical predictions.

Thus, the second SPP-2 processor was successful in obtaining new scientific results. Moreover, in order to substantiate the correctness of his work, important scientific results were obtained in the field of using random numbers in modeling problems in statistical physics.

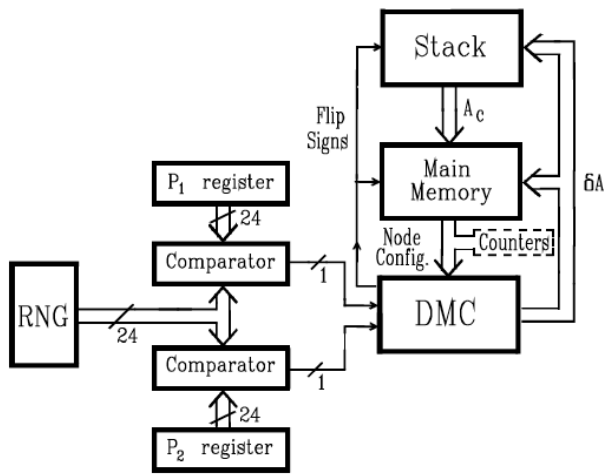

FIG. 1. Block diagram of the cluster processor SPP-2.

Figure 1 shows the basic schematic of an SPP-2 cluster processor. The RNG block is a programmable random number generator, the P1 and P2 registers contain the probabilities of inclusion in the cluster, the DMC (Decision Making Circuit) block decides whether the spin is included in the cluster, the Stack contains the raw spins of the cluster. The average construction time for one Wolff cluster took 400 nanoseconds.

## IV. CLUSTER PROCESSORS FOR THREE-DIMENSIONAL ISING MODELS

The success with the first cluster processor SPP-2 allowed us to move on to the development of the cluster processor SPP-3 to study the three-dimensional Ising model [21]. This processor was developed to study a model with linear dimensions up to 256. It implemented the ideas of our theory of random number generators, which were programmable with a shift register size up to 16384. Each processor had two random number generators operating in parallel and programmed with different characteristic pairs of Mersenian primes. 12 processors were manufactured and placed in four control nodes. As a result of many years of operation of the SPP-3 complex, numerical data were obtained, which, as a result of processing, led to record estimates of the critical parameters and critical temperature of the three-dimensional Ising model [22]. Only 20 years later, a team from the United States managed to come close to our results through calculations on a specially built private cluster farm using conventional computing architecture [23].

Figure 2 shows a photograph of one of the twelve SPP-3 processors. On the right is the RNG dual programmable random number generator [24]. One generator is visible, the second is located parallel to the first, their result, after applying a bitwise exclusive OR, is sent to the comparator for comparison with a probability that depends on the ratio of the coupling constant to the temperature. The bottom block is the main static 11-nanosecond memory containing 16 million spins. The middle block is the block for implementing the one-cluster Wolff algorithm. On the right is a block for connecting power and communication with the control computer. A total of about 400 microcircuits are used in one processor. The work от SPP-3 was supported by grants 07-13-210 NWO (Netherlands), INTAS-93-0211, M0Q000 ISF, and 93-02-2018 RFBR.

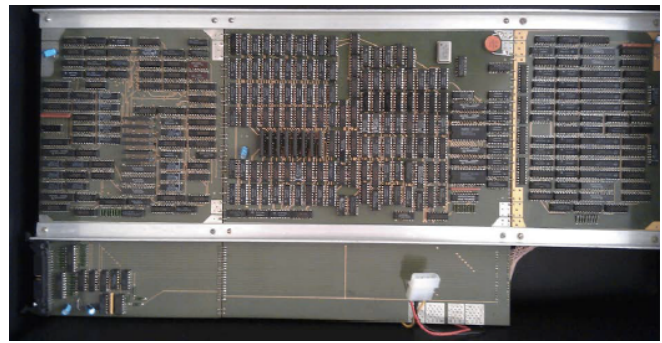

FIG. 2. Special-purpose processor SPP-3.

## V. CONCLUSIONS

1. Using specialized processors is much cheaper than supercomputers.
2. The results obtained with the help of specialized processors are chronologically ahead of the possibilities of supercomputer use.
3. Obtaining new results with the help of specialized processors is risky - in a large number of projects, performers have failed to achieve new scientific results.
4. There is a great risk due to errors or the discovery of new patterns since the accuracy of the calculations exceeds those preliminary estimates that were made using conventional computing technology.
5. The cost of human labor is very high.
6. Obtaining a new scientific result requires several years of work.
7. If successful, it is actually success!